\documentclass[rmp,twocolumn]{revtex4}

\usepackage{natbib}
\usepackage{times}

\usepackage{amsmath, graphicx, amssymb,tabls}
\usepackage{psfrag}

\newcommand{\beq}{\begin{eqnarray}}
\newcommand{\eeq}{\end{eqnarray}}
\newcommand{\beqn}{\begin{equation}}
\newcommand{\eeqn}{\end{equation}}

\newcommand{\bal}[1]{\begin{align} #1 \end{align}}

\newcommand{\expt}[1]{\langle #1 \rangle }

\newcommand{\pp}[1]{\left( #1 \right) }
\newcommand{\ra}{\rightarrow}

\newcommand{\ee}{\text{e}}

\newcommand{\BrB}{{\text{B}\rightarrow\text{B}}}
\newcommand{\Brb}{{\text{B}\rightarrow\text{b}}}
\newcommand{\brB}{{\text{b}\rightarrow\text{B}}}
\newcommand{\brb}{{\text{b}\rightarrow\text{b}}}

\begin{document}

\title{An accurate model for genetic hitch-hiking}

\author{A. Eriksson$^{*}$, P. Fernstr\"om$^{\dagger}$, B. Mehlig$^{\dagger}$, and S. Sagitov$^{\ddagger}$\\
$^{*}$\emph{\small Department of Energy and Environment, Chalmers University of Technology, G\"oteborg, Sweden}\\
$^{\dagger}$\emph{\small Department of Physics, G\"oteborg University, G\"oteborg, Sweden}\\
$^{\ddagger}$\emph{\small Mathematical Sciences, Chalmers University of Technology/G\"oteborg University, G\"oteborg, Sweden}}

\date{\today}

\begin{abstract}
We suggest a simple deterministic approximation for the growth of the favoured-allele frequency during a selective sweep. 
Using this approximation we introduce an accurate model for genetic 
hitch-hiking.  Only when $Ns < 10$  ($N$ is the population size and
$s$ denotes the selection coefficient), are discrepancies between our
approximation and direct numerical simulations of a Moran  model noticeable.
Our model describes the gene genealogies of a contiguous 
segment of neutral loci close to the selected one,
and it does not assume that the selective sweep happens instantaneously.  This enables us to compute SNP distributions on the neutral segment without bias.
\end{abstract}

\maketitle

%\tableofcontents
%\newpage

\section{Introduction}
\label{sec:introduction}

Gene genealogies under neutral evolution are commonly described
by the so-called coalescent process
\citep{kin82:coa,hudson83,hudson90,nor01:coa,hudson02}, 
incorporating recombination, geographical and demographical structure.
An important question is how gene genealogies are modified by
deviations from neutrality due to positive selection.
The answer to this question would help understanding 
to what extent and in which way selection has shaped the empirically
observed patterns of genetic variation.

Many  authors have addressed this question by 
considering the effect of a selective sweep at a given
locus on the gene history at a neighbouring
neutral locus. The dynamics of the selective sweep itself has been modelled in different ways.  Most
commonly, a deterministic model of the dynamics of the favoured-allele
frequency has been adopted
\citep{stephan_etal92,bra95:hit, kim_stephan02, przeworski02}, 
a notable exception 
being the early work of \citet{kap89:hit}.
Any deterministic model is of course an approximation to a more
appropriate model, such as  Moran or Wright-Fisher models of directional selection, 
where the allele frequencies fluctuate randomly in time.  The reasons
for attempting to ignore 
these fluctuations  are practical ones: 
the exact simulations are very time consuming \citep{kap89:hit}, and, in addition, deterministic models are much more amenable to theoretical analysis than the stochastic models.

Recently, \citet{durrett_schweinsberg04} have discovered an elegant 
asymptotic model (referred to as  the `DS-algorithm' in the following) 
for the genealogy of a single neutral locus during a selective sweep occurring 
in  its   vicinity. 
As the population size $N$ tends to infinity, 
their coalescent process approximates the Moran model \cite{moran58}
with recombination and positive selection.
\citet{durrett_schweinsberg04} have argued that
the fluctuations of the favoured-allele frequency during a selective sweep
may have a significant effect
on the gene-genealogy of a neighbouring
neutral locus, and hence on the distribution
of single-nucleotide polymorphisms (SNPs) at that locus. 
In a range of parameters determined by \citet{durrett_schweinsberg04}, 
the DS-algorithm describes the effect of a selective
sweep on the gene genealogy of a neutral locus nearby
very accurately, in close agreement with numerical simulations of a Moran model.

In this paper we suggest an efficient alternative to the DS-algorithm which is
equally accurate for the parameters
considered in \cite{durrett_schweinsberg04}, as shown in Fig. 
\ref{fig:part_distr_a}. 
For practical purposes, our algorithm has a number of advantages.
First, it allows for SNPs to occur during the
selective sweep because we do not assume that
the sweep happens instantaneously as does the
paint-box construction \cite{schweinsberg_durrett05}.  This avoids
a bias in the patterns of genetic variation at
the neutral loci  when the number of lines
in the sweep is not untypically small. 
Second, in practical applications, the question
usually is how selection affects genetic variation
in a contiguous stretch of neutral loci, whereas
the DS-algorithm describes the gene genealogy of
a single locus. Our algorithm, by contrast, determines
the ancestral recombination graph of an entire segment
of neutral loci close to a selected one. For example,
Fig. \ref{fig:part_distr_a} was obtained by a single run
of our algorithm. Third, our new algorithm gives an
accurate description of selective sweeps in
a much wider parameter range than the algorithm
proposed by \citet{durrett_schweinsberg04}.

On the theoretical side,  we propose an
efficient and accurate method for averaging
over the fluctuations of the  favoured-allele frequency.
Our scheme gives rise to a deterministic approximation
to the time-dependence of the favoured-allele frequency
during the sweep which, however, is very different
from the commonly used logistic model. Our model
is as easily implemented as the logistic model, but
much more accurate: it gives a very good
description of the genealogy of contiguous stretch
close to a selected locus provided $N s > 10$ where
$s$ parametrises the selective advantage of the favoured allele.
By contrast, the DS-algorithm  \cite{schweinsberg_durrett05}
requires $r\log(2N)/s \lesssim 1$ in order to be accurate, where $r$ is the recombination rate per individual per generation between the selected and the 
neutral locus. The logistic model requires very strong selection 
and large population size (see Figs. 8-10).

The remainder of this paper is organised as follows.
In section \ref{sec:selective_sweeps}, we give a brief account of previous
models of selective sweeps and their influence on the genealogies
of nearby loci (usually referred to as \lq genetic hitch-hiking', see
below). In section~\ref{sec:moran}, we describe our implementation 
of the Moran model. As \citet{durrett_schweinsberg04} we employ
Moran-model simulations as a benchmark for our new algorithm.
This new algorithm rests on two parts: a deterministic
model for the favoured-allele frequency during the sweep
(described in section~\ref{sec:formulation_problem}) and
the coalescent process
for a contiguous segment of neutral loci on the same chromosome 
as the selected locus (section~\ref{sec:bg_coalescent}). 
In section~\ref{sec:results}, we summarise our results, and conclude
in section~\ref{sec:conclusions}.

\section{Selective sweeps and genetic hitch-hiking}
\label{sec:selective_sweeps}
\subsection{Selective sweeps}

Consider the genetic composition at a certain locus in a diploid population with a constant generation size $N$. Suppose all $2N$ gene copies were of the same form b when a new allele B appeared due to a beneficial 
mutation. Let the new allele B have a 
fitness advantage (parametrised by $s$) 
as compared to the wild-type allele b. The frequency $x(t)$ of 
allele B at time $t$ is a stochastic process which 
exhibits a tendency to grow, but which may also become fixed at $x = 0$ 
(due to genetic drift) corresponding to the extinction of allele B. 
Once $x(t)$ has grown sufficiently from the initial low value $x(0)=1/2N$, 
the probability of reaching $x=1$ is high; eventually B takes over the
population. This process is usually referred to as a \lq selective
sweep'. In the limit of infinite population size, a selective
sweep is well approximated by the deterministic model
\begin{equation}\label{eq:logistic}
  \frac{{\rm d} x}{{\rm d}t} = s\,x (1 - x) ,
\end{equation}
see \citet{durrett_schweinsberg04} and the references cited therein.
Eq.~(\ref{eq:logistic}) is called the \lq logistic-growth equation'.

This growth model is a deterministic
approximation to the stochastic growth of $x(t)$. 
The latter is usually modeled in terms of
the Wright-Fisher model
\citep{wri31:evo,fis30:gen} with directional selection. 
This is a haploid population model with 
non-overlapping generations where reproduction is  
described by a biased sampling 
procedure with replacement: chromosomes are sampled
randomly, with replacement, from 
the previous generation, so that the ratio of the
probabilities of choosing a chromosome 
with the favoured allele to that without the favoured allele is $1:(1\!-\!s)$.
Direct numerical simulations of the Wright-Fisher model are commonly
employed to determine strengths and weaknesses of deterministic
approximations such as eq.~(\ref{eq:logistic}).

In the following we do not employ the Wright-Fisher model
as a reference, but a closely related model
with overlapping generations introduced by \citet{moran58}. 
As shown by  \citet{etheridge_etal06} it approximates the 
Wright-Fisher model when the population size is large. 

\subsection{Genetic hitch-hiking}

Consider the genetic variation at a neutral locus on the same 
chromosome as the selected locus. Clearly, the pattern of
genetic variation at the neutral locus is  
influenced by a selective sweep in its vicinity -- the smaller the distance 
the larger the influence. When the B allele first appeared in
the population  because of a favourable mutation, the 
corresponding alleles at the neutral locus have more 
offspring compared with other alleles not associated with the 
B allele on the selected locus. Thus, the favoured alleles at the neutral 
locus are spread through the population to a larger extent than can be 
explained in a neutral model. This effect is known as genetic hitch-hiking 
\citep{may74:hit}. Far from the selected locus, recombination will 
effectively eliminate linkage between the neutral and selected loci, so 
that the influence of the selective sweep becomes negligible.

\begin{figure}
\centerline{
\psfrag{AA}{(i)}
\psfrag{BB}{(ii)}
\psfrag{CC}{(iii)}
	%\framebox[8cm]{\raisebox{5cm}}
	\includegraphics[width=8cm]{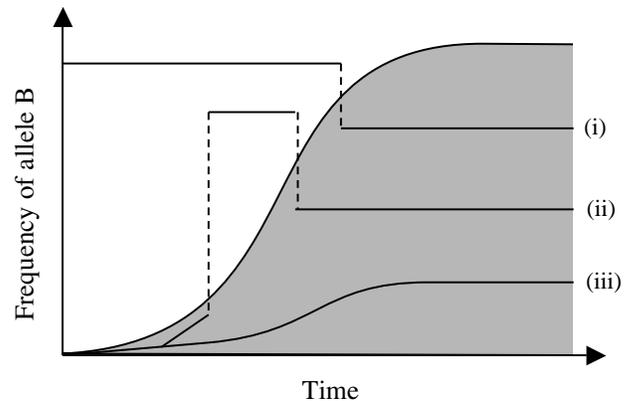}
}
\caption{\label{fig:escaped_neutral_lines}
Illustration of the hitch-hiking effect on the ancestral lines of a neutral locus. The shaded area corresponds to 
individuals with the advantageous allele B at the selected locus in the population. Close to the selected locus, most lines are identical by descent to the originator of the sweep (line (iii)). Recombination 
(shown as dashed lines) can cause a line to escape the sweep, i.e. the originator does not belong the 
ancestral line, because at some stage a recombination event causes the allele at the neutral locus to be inherited from an ancestral line that has not yet been caught the sweep (line (i)). Much less likely, but still possible, is for the line to first escape but later recombine back into the path of the sweep (line (ii)). After \citet{durrett_schweinsberg04}.
}
\end{figure}

Figure~\ref{fig:escaped_neutral_lines} illustrates the hitch-hiking 
effect in terms of the ancestral graph for a small hypothetical sample of sequences 
taken at a neutral locus. (For the sake of clarity we assume that the 
selected locus is located left of the neutral locus of 
interest.)  Most ancestral lines can be traced back to the originator of 
the sweep, but some lines exhibit recombination events allowing them to 
escape from the sub-population with the B allele.

It is straightforward but cumbersome to directly simulate the Wright-Fisher
(or Moran) model in order to analyse how patterns of
genetic variation are affected by hitch-hiking.
Several authors have therefore studied approximations to the growth process of the selected allele frequency $x(t)$. \citet{kap89:hit} divide the selective sweep into three phases: the early phase is modeled by a supercritical branching process, the middle phase is described by the deterministic logistic growth, and the final phase is viewed as a sub-critical branching process. The probability that the sweep succeeds is
approximately given by the selective advantage $s$, when $s$ is small. As a
consequence, one may need
to iterate this procedure many times to collect enough successful simulations. 

This approach has been simplified by ignoring the
initial and final (stochastic) phases \citep[see, e.g.,][]{stephan_etal92, bra95:hit, kim_stephan02, przeworski02} and
instead using
the deterministic logistic model (\ref{eq:logistic}) for the whole sweep. 
This makes it possible to simulate the sweep backwards in time, which
in turn enables one to perform computations conditional
on that the sweep succeeds. This approach is significantly
faster than an algorithm based on the better approximation by
\citet{kap89:hit}.

\citet{barton98} \citep[see also][]{per97:evo} has considered a stochastic
shift between the introduction of the favoured allele and the onset of
the deterministic growth; the distribution of the shift is derived from
modelling the spread of the beneficial allele in the initial phase of
the sweep as a super-critical branching process. This approximation
captures some of the effects of the conditioning on the success of the
sweep and the stochastic growth in the early stages of the sweep. The
middle and late stages of the sweep are treated in the logistic approximation. 
Within his model, Barton gives analytical expressions for
the probability that two copies of a neutral marker are identical by
descent, assuming that any recombination event leads to ancestral lines
escaping the sweep.

As argued by \citet{durrett_schweinsberg04}, 
the disadvantage of ignoring the fluctuations 
is that the probabilities of how lines merge and recombine 
are not correctly described.  
They consider the gene genealogy of a selected locus 
and a nearby neutral locus and propose 
an elegant approximation to the Moran dynamics, valid
in the limit of large population size and strong selection, 
which  captures the stochastic aspects of the sweep,
and correctly models the partitioning of the neutral lines
as a consequence of the selective sweep.

\section{The Moran model of positive selection}
\label{sec:moran}

In this section we describe the Moran model \cite{moran58} 
for the evolution of  
a diploid population of $N$ individuals.  The Moran model
is used as a benchmark to test the accuracy of our coalescent
model described in sections \ref{sec:formulation_problem}
and \ref{sec:bg_coalescent}.

We consider a chromosome
with a locus subject to positive selection and determine  both the evolution of this selected locus,
as well as  genealogies of neutral loci in its vicinity. In the first
subsection we describe the growth of the favoured-allele frequency in the population. 
In the second subsection we explain how to condition this process on 
the success of the selective sweep. This is necessary because in trying
to deduce the effect of a sweep on neutral loci nearby we assume that
the sweep actually took place. In the last subsection we summarise
how gene genealogies of such neutral loci are calculated
within the Moran model.

\subsection{Spread of the advantageous allele during the sweep}

As in the previous section we assume that
there is a favoured allele at the selected locus, B say, 
and a set of selectively neutral variants, which we will refer to collectively as b.  
The life-time of each individual is taken to be an independent exponentially distributed variable with  expected value of one generation. 
When an individual dies, it is replaced with a copy of an individual chosen with replacement with uniform probability from the whole population, 
except that replacement of an individual with the B allele with an individual with the b allele is rejected with probability $s$; 
this is what constitutes selection in this model. Instead, a parent is chosen with uniform probability from the set of individuals with the B allele. 
Thus, $s=0$ corresponds to neutral evolution and $s=1$ is the strongest possible selection. In short, 
the population evolves according to a time-continuous Markov process where the different 
events occur with rates
\begin{align}\label{eq:moran_rates}
	w_\brb =&\  2N \times \left( 1 - \frac{k}{2N} \right) 
			 \times \left( 1 - \frac{k}{2N} \right) \,,\nonumber \\
	w_\brB =&\  2N \times \frac{k}{2N} \times \left(1-\frac{k}{2N}\right)\,, \nonumber \\
	w_\Brb =&\  2N \times \frac{k}{2N} \times \left(1-\frac{k}{2N}\right)\left(1-s\right)\,, \nonumber \\
\nonumber	w_\BrB =&\  2N \times \frac{k}{2N} \times \frac{k}{2N}\ +  \\&\ 
	               2N \times \frac{k}{2N} \times \left(1-\frac{k}{2N}\right) s\,.
\end{align}
The three factors in the rates $w_{\alpha \ra \beta}$, where $\alpha$ and $\beta$ stand for
either b or B, have the following interpretations:
The first factor is the total rate of replacement events in the population per generation;
the second factor is the probability that the line that dies has the allelic type $\alpha$; the final factor is the probability that the replacing line has the allelic type $\beta$. The second term in the rate $w_\BrB$ corresponds to the rejected B-to-b replacements. It follows from eq. (\ref{eq:moran_rates}) 
that the sum of events is $2N$ per generation for all values of $s$.

\citet{durrett_schweinsberg04} use a slightly different version
of the Moran model with positive selection. In their model, the rejected B-to-b transitions are ignored, 
whereas we take them to be B-to-B transitions.
This difference does not affect the trajectory of the number of copies of the advantageous allelic type. 
A third possibility would be to introduce selection by means of a
probability of survival to maturity which would be $1$ for B alleles, but
$1-s$ for b alleles. The corresponding  modifications of eq.
(\ref{eq:moran_rates}) would require minor changes to the background
coalescent described in section \ref{sec:bg_coalescent},
but we do not discuss these here.

\subsection{Conditioning on the fixation of allele B}

In each replacement, the number of copies $k$ of allele B in the population is either increased by one (corresponding to a $\brB$ event), decreased 
by one (corresponding to a $\Brb$ event), or left 
unchanged (corresponding to a $\BrB$ or $\brb$ event). Consider the number $k_i$ of copies of the advantageous allelic type in the population after the $i$th change in $k$. The sequence $k_1, k_2, \ldots$ then follows a Markov chain, where the probability that $k$ is increased by one after a replacement where $k$ changes is 
\bal{\label{eq:k_increase}
	\frac{w_\brB}{w_\brB + w_\Brb} = \frac{1}{2-s}.
}
The probability $h_{k}$ of fixation of the B allele in the population, given that there are $k$ copies at present, equals the probability of fixation after a change in $k$. With the probability that $k$ increases in (\ref{eq:k_increase}), one obtains the recursion
\beqn
	h_{k} = \frac{1}{2-s}\, h_{k+1} + \Big(1 - \frac{1}{2-s}\Big) h_{k-1}
\eeqn
where $k$ is between $1$ and $2N-1$. If $k$ is zero, there are no copies of B that can reproduce; hence, $h_0 = 0$. Similarly, when $k = 2N$ all individuals in the population has the B allele, corresponding to $h_{2N}=1$. With these two conditions the recursion has a unique solution, given by \citep[see, e.g.,][and references therein]{durrett02} 
\beqn
	h_{k}=\frac{1-\left(1-s\right)^{k}}{1-\left(1-s\right)^{2N}}.
\eeqn
Usually, the population size is large and the selection parameter is small. If in addition $2Ns$ is large, we obtain the well-known result that the probability $h_{1}$ that the sweep succeeds from a single copy of the B allele is approximately $s$.
This means that if the sweep is initiated with a single copy of the B allele, and the rates are given  by (\ref{eq:moran_rates}), in most cases the B allele will become extinct in a few generations because of the fluctuations in the early stage of the sweep. When $k$ reaches a critical level (where $ks$ is relatively large), the probability that the fluctuations will cause B to become extinct becomes exponentially small; thus, a sweep that escapes this level will almost certainly continue to increase in abundance and eventually become fixed in the population.

In this paper, we consider only sweeps that succeed. It is thus necessary to consider the Markov chain conditioned on the success of the sweep. The conditioning does not change the
rate of events replacing 
an individual for one of the same kind, since they do not affect the success of the sweep. 
The new rates become \citep{durrett_schweinsberg04}:
\bal{\label{eq:tilde_w}
	\widetilde{w}_\brB(k)
	   &= w_\brB(k) \, \frac{h_{k+1}}{h_k}  = \frac{k\left(2N-k\right)}{2N} \frac{1-\omega^{k+1}}{1-\omega^k}\,,\\
	\widetilde{w}_\Brb(k)
		 &= w_\brB(k) \, \frac{h_{k-1}}{h_k}= \frac{k\left(2N-k\right)}{2N} \frac{\omega-\omega^{k}}{1-\omega^k}\,,
                 \nonumber\\
	\widetilde{w}_\BrB(k) &= {w}_\BrB(k)\,,\nonumber\\
	\widetilde{w}_\brb(k) &= {w}_\brb(k)\nonumber\,.
}
where $\omega = 1-s$.
Thus, we can simulate the embedded Markov chain of the changes in $k$, conditioned on the success of the sweep if we take the probability $p_{+}(k)$ of going from $k$ to $k+1$ copies of the B allele as
\beq
	p_{+}(k) &=& \frac{\widetilde{w}_\brB}{\widetilde{w}_\brB + \widetilde{w}_\Brb} 
        = \frac{1-\omega^{k+1}}{(1+\omega)(1-\omega^k)} \label{eq:p_plus_k}\,.
\eeq
The probability that the number of alleles decreases from $k$ to $k-1$ is $p_{-}(k) = 1  - p_{+}(k)$.

\begin{figure}
\psfrag{xlabel}[t][]{Time}
\psfrag{ylabel}[][]{Number of copies of B}
\psfrag{xlabel2}[][]{Time}
\psfrag{ylabel2}[][]{Copies of B}
\centerline{\includegraphics[width=8.5cm]{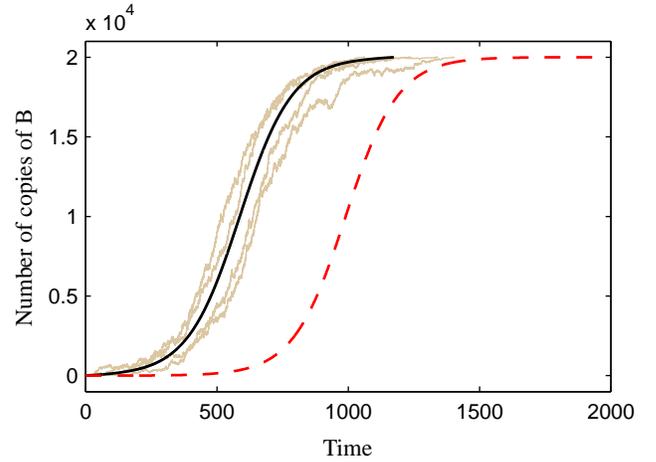}}
\caption{\label{fig:moran_example}
Growth of the favoured-allele frequency in the population (time is measured in generations). The population size is $N=10^4$, 
and the selection parameter is $s=0.01$. Shown are four samples of the Moran process (grey lines), the logistic model (dashed red line), and our new deterministic model described in section
\ref{sec:det_model}, solid black line. 
The new deterministic approximation (\ref{eq:exp_T_n_sol}) is much closer to the Moran curves than the logistic approximation.  }
\end{figure}

Fig. \ref{fig:moran_example} shows four realisations of 
the favoured-allele frequency  
$x(t)$ generated with the algorithm described
above. Also shown is the logistic model for $x(t)$ (dashed line)
which is not a good approximation, as well as our new model
described in section \ref{sec:formulation_problem}, solid line.

\subsection{Gene genealogies of the neutral loci during the sweep}

In this section, we describe our implementation of the Moran model for simulating the 
gene genealogies of neutral loci in the neighbourhood of a  selected locus. 
The algorithm is divided into a forward and a backward phase. 

In the forward phase, we generate the sequence of the number $k$ of B alleles, forward in time, according to the conditioned Markov process described in the previous section:
starting from $k=1$, $k$ is incremented with probability $p_{+}(k)$, 
or decremented with probability $1 - p_{+}(k)$, until $k = 2N$. Because we either increase or decrease $k$, each value in the sequence is different from the previous one.

In the backward phase, the population is divided into two sub-populations with B or b alleles at the selected locus.
At the end of the sweep, all ancestral lines are in the B population; this is the starting point for the backward phase.
We trace the genealogies of the neutral loci backward in time by 
traversing the sequence of $k$ values (obtained in the forward pass) in reverse; this guarantees that the time-reversal of the Moran process is correct. Each time $k$ changes, we generate a $\brB$ event if the new value of $k$ is smaller than the old one. Correspondingly, we  generate a $\Brb$ event if $k$ increases.  Between each change in $k$, we 
generate the $\BrB$ and $\brb$ events of the Moran chain (these events does not change $k$). 
The number $m$ of such events has a geometric distribution, $q_k\left(1-q_k\right)^m$, where
\beqn\label{eq:q_val}
	q_k = (2-s)\frac{k}{2N}\left(1 - \frac{k}{2N}\right).
\eeqn
The probability that the event is a $\brb$ replacement is 
\bal{
	\frac{\widetilde{w}_\brb}{\widetilde{w}_\BrB + \widetilde{w}_\brb} = \frac{(2N-k)^2}{\pp{2N}^2 - \pp{2-s}k\pp{2N-k}},
} 
and, correspondingly, the $\BrB$ replacements occur  with probability 
$\widetilde{w}_\BrB/(\widetilde{w}_\BrB + \widetilde{w}_\brb)$.
Finally, the time between each event is exponentially distributed with expected value $(2N)^{-1}$ in units of generations. 

We now describe the effect of the events generated during the sweep on the gene genealogies of the neutral loci.
In each event, we choose the line to die and the line to replace it randomly from the appropriate sub-populations. 
As we proceed backward in time, the dying line  coalesces with its parent line (e.g, in a $\Brb$ event, we pick the line to coalesce from the b sub-population). 
With probability $r$, recombination occurs between the selected locus and the right-most locus during the coalescent. In this case,
the region between the selected locus and the recombination point coalesces with the chosen parent, and the second part of the neutral region, between the recombination point and the rightmost locus, coalesces with a parent chosen with uniform probability from the whole population.
We assume that the neutral locus of interest is sufficiently small so that
there is at most one crossover event in the region in each meiosis 
(the deterministic coalescent models, however, are not subject to this limitation since in these models the recombination rate can be arbitrarily high).
For the values of $r$ considered in this article this approximation is good. 
If necessary, it is straightforward to improve it, 
for instance by simulating an explicit recombination process instead of simply assuming that no or one crossovers occur in the interval in each meiosis. One may also implement more realistic models of recombination, e.g. models 
which capture crossover interference \citep[see, e.g.,][for a review]{mcpeek_speed95}; for the purpose of this paper, however, the simplest 
model is sufficient.

When the simulation has reached the beginning of the sweep, there is exactly one line carrying the B allele, and the genetic material of this individual is ancestral to all genetic material trapped in the sweep. 
In addition, there may be a set of lines which have escaped the sweep because of recombination as explained in section~\ref{sec:selective_sweeps}.
We then follow the lines carrying genetical material from the sample back in time until the most recent common ancestor of each locus has been found for the sample. Since there is no selection in this part of the history, the Moran process is a coalescent where the rate (in units of events per generation) of two lines coalescing is $n(n-1)/2N$, where $n$ is the number of lines in the population, and the rate of recombination is $r$.

\section{Averaging over realisations of the sweep}
\label{sec:formulation_problem}

\citet{durrett_schweinsberg04} have convincingly shown  that it
is necessary to consider the fluctuations of the favoured-allele
frequency (displayed in Fig. \ref{fig:moran_example}) in order
to accurately represent effects of the sweep on nearby
loci. 

We now explain how to efficiently and accurately
average over such fluctuations. We motivate our method
by an example: how to compute the probability 
that the first recombination event, if it occurs during the
sweep, occurs with an individual not carrying the favoured allele at
the selected locus.
In section \ref{sec:bg_coalescent} we describe a coalescent process
which makes use of the ideas described in this section.

\subsection{An example}
We illustrate our approach by considering the conditional 
probability $Q(r)$ that the first recombination event, if it occurs during 
the sweep, occurs with an individual not carrying  the favoured allele at the selected locus: 
\bal{\label{eq:Q_def}
	Q(r) = \int_0^{\tau} \!{\rm d}t\,r\,e^{-rt}\, [1 - x(t)]  \,.
}
$Q(r)$ depends on the realisation of $x(t)$ of the sweep of duration $\tau$. For small values of
$r$, it is unlikely that a given line experiences more than one recombination event 
during the sweep, and in this case $Q(r)$ is approximately the probability that the line escapes the sweep.

\begin{figure}
\centerline{
\psfrag{x}[][]{$r$}
\psfrag{y}[][t]{$\expt{Q(r)}$}
	\includegraphics[width=8.5cm]{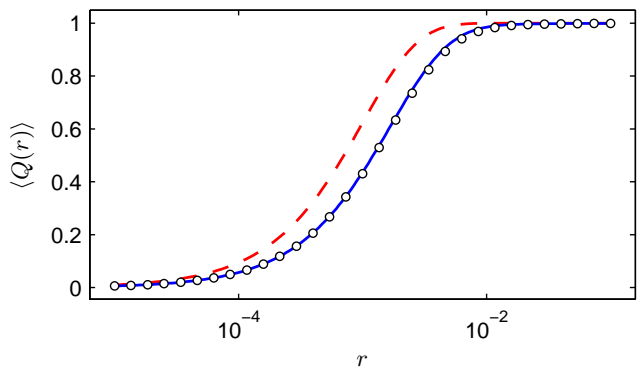}
}
\caption{\label{fig:q_compare}
Comparison of $\expt{Q(r)}$ as a function of $r$ for the different models: Moran simulations (circles),
the deterministic logistic model (dashed red line), and the new deterministic model (solid blue line). The population size is $N=10^4$ and the selection parameter is $s=0.01$. 
}
\end{figure}

Fig. \ref{fig:q_compare} shows the average $\langle Q(r)\rangle$ over realisations
of $x(t)$ as a function of $r$,
obtained from Moran-model simulations (circles). Also shown
are the results from the logistic model (dashed line), derived   
as follows. Inserting the solution of (\ref{eq:logistic})
\bal{
   x(t) = \frac{1}{1 + \ee^{-s\pp{t - \tau/2}}}
}
(where  $\tau = 2\ln(2N-1)/s$ is the duration of the sweep in the
logistic model),
into (\ref{eq:Q_def}) and expanding the integrand in (\ref{eq:Q_def}), we obtain
\bal{\label{eq:q_logistic}
   \langle Q(r) \rangle= 1 \!- \!\ee^{-r\tau/2} \! + \!
   \sum_{n=1}^\infty (-1)^n \, 2r^2\, \frac{\ee^{-r\tau/2} \!-\! \ee^{-ns\tau/2}}{s^2 n^2 - r^2}\,.
}
As can be seen in Fig. \ref{fig:q_compare}, the result (\ref{eq:q_logistic}) deviates
significantly from the Moran-model results.

We now show how to obtain a much more accurate approximation (solid line
in Fig.~\ref{fig:q_compare}). 

The problem in averaging (\ref{eq:Q_def}) over different realisations of the stochastic Moran sweep 
lies in that both the upper bound $\tau$ of the integral and the integrand fluctuate.
In the following we describe an approximate method of averaging (\ref{eq:Q_def}) 
which gives accurate results and motivates a new deterministic model
for selective sweeps.
To begin with, note that $x(t)$ is piecewise constant function of time in
the Moran model. A realisation of the growth of the B allele is determined
by a sequence of $M$ pairs $(k_i,\tau_i)$ 
where $k_i$ is the number of copies of B in time interval $i$, and 
$\tau_i$ is the duration of this interval (the latter
begins at $t_i = \sum_{j=1}^{i-1}\tau_j$).
The sweep begins with $k_1 = 1$ at time $t_1 = 0$, and ends with $k_M = 2N$ at time $t_M$. 
Thus, we have
\begin{equation}
\label{eq:Q2}
  Q(r) =  \sum_{i=1}^{M-1} \left[ \ee^{-r \, t_i} - \ee^{-r \,t_{i+1}} \right] \frac{2N-k_i}{2N}\,.
\end{equation}
The number $M$ of steps in the growth process fluctuates and is usually much greater
than $2N-1$ since $k_i$ is usually not an increasing function of $i$.

\begin{figure}
\centerline{
\psfrag{x}[][]{Time}
\psfrag{y}[][t]{No. of copies of B}
	\includegraphics[width=8.5cm]{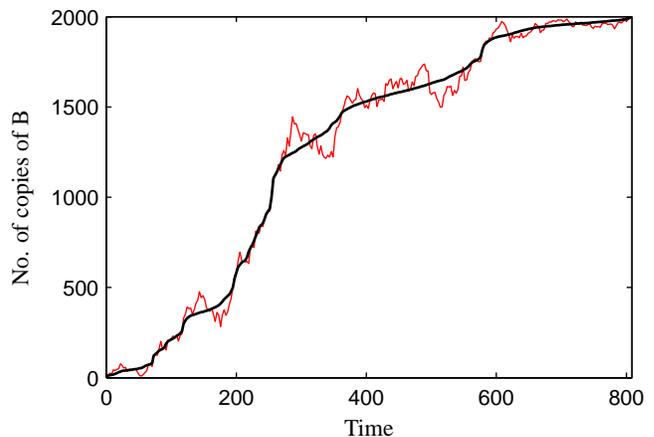}
}
\caption{\label{fig:shuffle}
Comparison between the actual growth curve $k_i$ versus $t_i$, red line, and 
the corresponding sorted curve $\widetilde{k}_i = i$ versus $\widetilde {t}_i$, black line. 
The parameters are $N = 10^3$ and $s=0.01$. 
}
\end{figure}

We construct an increasing growth curve from the sequence $(k_i,\tau_i)$ as follows. 
First, consider the sequence obtained by sorting the intervals such that $k_i \le k_{i+1}$. 
Second, merging all intervals with the same value of $k_i$ 
into one contiguous segment, we obtain a sequence of $2N-1$ segments, 
$(\widetilde{k}_i = i,\widetilde{\tau}_i = \sum_{j:k_j=i} \tau_j)$,  with
$\widetilde{t}_i = \sum_{j=1}^{i-1} {\widetilde\tau}_j$
so that $\widetilde t_{2N}$ is the duration of the sweep.
Note that $\widetilde{t}_i$ may also be written as $\sum_{j:k_j<i} \tau_j$, which implies $\widetilde{t}_{2N} = t_{2N}$.
This \lq sorted' sweep is monotonous:
there are $i$ copies of allele B in the population during the 
time interval $[\widetilde{t}_i, \widetilde{t}_{i+1}]$, 
and at time $\widetilde{t}_{i+1}$ the number of copies of B increases by one. 
Fig.~\ref{fig:shuffle} shows that this results in a 
surprisingly accurate representation of the original trajectory $x(t)$.
This is so because of the conditioning on the success of the sweep: 
large downwards fluctuations of $k_i$ are rare.

In terms of the \lq sorted' sweep, eq. (\ref{eq:Q2}) can be written
as
\begin{equation}
\label{eq:Q3}
  Q(r) \approx  \sum_{k=1}^{2N-1} \left[ \ee^{-r \, \widetilde{t}_k} - \ee^{-r \,\widetilde{t}_{k+1}} \right] 
  \frac{2N-k}{2N}\,.
\end{equation}
Averaging (\ref{eq:Q3}) over the realisations of the sweep is straightforward. 
Assuming that $\langle \exp(-r \,\widetilde{t}_k)\rangle$
can be approximated by $\exp(-r \langle \widetilde{t}_k\rangle )$,
we find
\begin{equation} 
\label{eq:expt_q_5}
  \expt{Q(r)} \approx  
  				\sum_{k=1}^{2N-1} \left[ \ee^{-r \, \expt{\widetilde{t}_k}} - \ee^{-r \,\expt{\widetilde{t}_{k+1}}} \right] 
  			   \frac{2N-k}{2N}\,.
\end{equation}
The expectation values $\langle \widetilde t_k\rangle$ can be calculated
analytically as shown in
section~\ref{sec:expected_value_T_n} below.  
In Fig.~\ref{fig:q_compare}, $\expt{Q(r)}$ according to  
(\ref{eq:expt_q_5}) is shown as a blue line,
in very good agreement with the numerical data (circles).

\subsection{A deterministic model for $x(t)$}
\label{sec:det_model}
Our result (\ref{eq:expt_q_5}) can be written in the form (\ref{eq:Q_def})
by introducing a deterministic model for the sweep. Let 
$\bar{k}(t)$ be the solution of $\langle \widetilde t_{{k}}\rangle=t$ for ${k}$.
In Fig.~\ref{fig:moran_example}, $\bar{k}(t)$ is shown as a solid black line.
Let $\bar{x}(t) = \bar{k}(t)/(2N)$. Then
\begin{equation}
   \langle Q(r) \rangle \approx \int_0^{\bar{\tau}} \!{\rm d}t \, r \,{\rm e}^ {-rt} \left[1-\bar{x}(t)\right]
\end{equation}
where $\bar{\tau} = \langle \widetilde{t}_{2N}\rangle$ is the expected duration of the sweep.

In practice, $\bar{k}(t)$ is obtained as follows: we pick $10^3$
linearly spaced values for $t$ in the interval $[0,
\expt{\widetilde{t}_{2N}}]$. For each value of $t$, we find the $k$
such that $\expt{\widetilde{t}_k} \le t \le \expt{\widetilde{t}_{k+1}}$,
using eq.~(\ref{eq:exp_T_n_sol}) to calculate the values of
$\expt{\widetilde{t}_k}$.
To find the value of $\bar{k}$ corresponding to $t$, we use linear
interpolation between the endpoints of this interval.

Results of coalescent processes based on the model $\bar{x}(t)$ for the selective
sweep are summarised in section~\ref{sec:results}.
As expected the results obtained exhibit equally good
agreement with our Moran-model simulation as does Fig.~\ref{fig:q_compare}.

\subsection{The expected value of $\widetilde{t}_k$}
\label{sec:expected_value_T_n}

In this section, we derive an analytical expression for 
$\expt{\widetilde t_k}$, 
the total time during the whole sweep when there are $k$ copies of B or less, starting from a single copy. 
More generally, let $T_i^{(k)}$ be the corresponding time, measured during the remaining parts of the 
sweep starting from $k$ copies of B. 
Thus, we have $\expt{\widetilde t_k} = \langle T_1^{(k-1)}\rangle$.
 
The value of $\expt{T_i^{(k)}}$ equals the expected time until the next event, plus the expected time spent in states with $k$ copies of B or less from the next state. Thus, we have the recursion
\beqn\label{eq:expt_T_recursion}
	\expt{T_i^{(k)}} = \expt{\tau_i} \, \theta_{k-i} + p_{+}(i) \expt{\tau_{i+1}^{(k)}} + p_{-}(i) \expt{\tau_{i-1}^{(k)}},
\eeqn
where $\theta_i$ is one if $i \ge 0$ and is zero else, and $p_{\pm}(i)$ is the probability of going from $i$ to $i\pm1$ copies of B,
c.f. Eq.~(\ref{eq:p_plus_k}). In order to find a unique solution to (\ref{eq:expt_T_recursion}), we need to provide boundary conditions. 
First, we note that 
the transition from $i=1$ to $i=0$ is forbidden (this is known as 
a \lq natural boundary condition'). 
Second, if the sweep is started at $i = 2N$ it stops immediately; thus, 
we must take 
\beqn
	\expt{T_{2N}^{(k)}} = 0
\eeqn
for all $k$. In the following it turns out to be convenient to introduce
\beqn\label{eq:def_phi}
	\phi_i^{(k)} = (1-\omega^i)\, \expt{T_i^{(k)}}.
\eeqn
Writing (\ref{eq:expt_T_recursion}) in terms of $\phi_i^{(k)}$ leads to a recursion with constant coefficients:
\beqn\label{eq:expt_phi_recursion}
	\phi_{i+1}^{(k)} - (1+\omega) \phi_i^{(k)} + \omega\,\phi_{i-1}^{(k)} 
	= - (1+\omega)(1-\omega^i) \expt{\tau_{i}}\,\theta_{k-i}.	
\eeqn
We solve (\ref{eq:expt_phi_recursion}) as follows. 
First, from (\ref{eq:expt_phi_recursion}) we obtain a recursion for the difference $\Delta_{i}^{(k)} = \phi_{i+1}^{(k)} - \phi_i^{(k)}$:
\beqn
	\Delta_{i}^{(k)} = \omega \Delta_{i-1}^{(k)} - (1+\omega)(1-\omega^i) \expt{\tau_{i}}\,\theta_{k-i}.
\eeqn
By telescoping from zero to $i$, we find the solution
\beqn\label{eq:Delta_sol}
	\Delta_{i}^{(k)} = \omega^{i} \Delta_{0}^{(k)} - 
						\sum_{j=1}^{i} \omega^{i-j} \pp{1-\omega^j}\pp{1+\omega} \expt{\tau_{j}}\,\theta_{k-j}\,.
\eeqn
At $i = 0$, (\ref{eq:def_phi}) implies $\phi^{(k)}_0 = 0$, which leads to $\Delta_0 = \phi^{(k)}_1$.
With this, summing (\ref{eq:Delta_sol}) from $0$ to $i-1$ leads to
\begin{align}\label{eq:expt_phi_recursion2}
	\expt{T_i^{(k)}} &= 
	\frac{1}{1-\omega^i} \sum_{j=0}^{i-1} \Delta_{j}^{(n)} \nonumber\\
		&= \expt{T_1^{(k)}} - 
		   \sum_{j=1}^{i-1}\frac{(1-\omega^{i-j})(1-\omega^j)}{(1-\omega^i)(1-\omega)} (1+\omega)\expt{\tau_j} \, \theta_{k-j}\,.
\end{align}
Setting $i = 2N$  in (\ref{eq:expt_phi_recursion2}), and using $\expt{T_{2N}^{(k)}} = 0$, we can solve for $\expt{T_1^{(k)}}$:
\begin{align}\label{eq:expt_phi_recursion3}
	\expt{T_1^{(k)}} =
	\sum_{j=1}^{k}\frac{(1-\omega^{2N-j})(1-\omega^j)}{(1-\omega^{2N})(1-\omega)} \left(1+\omega\right)\expt{\tau_j}\,.
\end{align}
Between each change in $k$, there is a geometrically distributed number of events. It follows from (\ref{eq:q_val}) that the expected time between two changes in $k$ is
\beq\label{eq:expt_tau}
	\expt{\tau_k} &=& \left[ \widetilde{w}_\brB + \widetilde{w}_\Brb \right]^{-1 } \nonumber \\
				     &=& 2N/\left[ k(2N - k)(1+\omega) \right] \,.
\eeq
generations. Inserting the value of $\expt{\tau_k}$ and writing the solution in terms of $\expt{\widetilde{t}_i}$, we obtain 
\beqn\label{eq:exp_T_n_sol}
	\expt{\widetilde{t}_k} = \sum_{i=1}^{k-1}\frac{2N\left(1-\omega^{2N-i}\right)\left(1-\omega^i\right)}{i\left(2N-i\right)\left(1-\omega\right)\left(1-\omega^{2N}\right)}\,.
\eeqn
Finally, we note that higher moments of $\widetilde{t}_k$, especially the variance, can be obtained in a similar manner. 

\section{The background coalescent for neutral loci in the vicinity of a
selected one}
\label{sec:bg_coalescent}

As explained in section~\ref{sec:selective_sweeps}, selection influences, 
via the 
hitch-hiking effect, the evolution of neutral loci on the same chromosome as the selected locus. 
Given a particular growth of the favourable allele frequency $x(t)$ as a function of time, what is the evolution of the linked neutral loci?

The standard approach is to follow \citet{kap89:hit} \citep[see also][]{kap88:coa} in modeling the effect of selection on the neutral loci as a form of population structure: The selective sweep is viewed  as a two-island population with migration, where one island, with population size $2Nx$, contains the individuals with the B allele; the other island has population size $2N(1-x)$ and
contains the individuals with the b allele. Coalescent events can occur only between individuals on the same island. Recombination, however, may move a line from one island to the other, since the parent of the second product of the recombination event is chosen uniformly from the whole population.

It is useful to write the total rate of coalescent and recombination events in the subdivided population in the form
\beqn\label{eq:lambda_tot_standard}
   \lambda_\text{tot} = \lambda_\text{B} \,p_\text{B} + \lambda_\text{b}\, p_\text{b},
\eeqn
where $\lambda_\text{B}$ and $\lambda_\text{b}$ are the total number of birth-death events per generation in the B and b sub-populations, respectively, is given by
\begin{align}\label{eq:lambda_standard}
  \lambda_\text{B} &= 2N \, x, \nonumber\\
  \lambda_\text{b} &= 2N \left(1 - x\right),
\end{align}
and where $p_\text{B}$ and $p_\text{b}$ are the probabilities that a single birth-death event leads to a coalescent or recombination event (or both) involving an individual in the corresponding sub-population.

Consider the probability $p_\text{B}$. First, a birth-death event has no effect on the gene genealogies 
unless the individual born is an ancestor to a locus of an individual in the sample. The probability that this is the case is simply $n_\text{B}/(2Nx)$, where $n_\text{B}$ is the number of ancestral lines currently in the B sub-population. Second, in order for the gene genealogies to change either recombination must happen during the birth -- this happens with probability $r$ -- or the parent must belong to a different ancestral line of the sample; the probability that this happens is $(n_\text{B}-1)/(2Nx)$.
Since one of the sub-populations can be quite small, especially close to the ends of the sweep, we cannot make the usual assumption \citep{hudson90} that recombination and coalescence cannot occur in the same event. Putting it all together, we find
\beqn\label{eq:p_B}
%   p_\text{B} = \frac{n_\text{B}}{2Nx}\,\frac{n_\text{B}-1}{2Nx}\left(1 - r\right) + n_\text{B} r.
 p_\text{B} = \frac{n_\text{B}}{2Nx}
    \left[ \left(1 - r\right) \frac{n_\text{B}-1}{2Nx} + r \right]\,.
\eeqn
The first term corresponds to two lines coalescing in the B population with no recombination, and the second term corresponds to all events involving recombination.

We derive the probability $p_\text{b}$ of an event in the b sub-population in the same way as for $p_\text{B}$. The result is
\beqn\label{eq:p_b}
%   p_\text{b} = \frac{n_\text{b}}{2N(1-x)}\,\frac{n_\text{b}-1}{2N(1-x)}\left(1 - r\right) + n_\text{b} r,
 p_\text{b} = \frac{n_\text{b}}{2N(1-x)}
    \left[ \left(1 - r\right) \frac{n_\text{b}-1}{2N(1-x)} + r
    \right]\,.
\eeqn
where, correspondingly, $n_\text{b}$ is the number of ancestral lines currently in the b sub-population.

When $x$ and the other parameters are constant, the coalescent is a Poisson process, and the time to the next event is exponentially 
distributed with expected value $1/\lambda_\text{tot}$, 
see Eq.~(\ref{eq:lambda_tot_standard}).
In a selective sweep, however, $x$ changes with time; hence, the coalescent is an inhomogeneous Poisson process. Given the state of the population at time $t_1$, the distribution $f(t_2|t_1)$ of the time $t_2$ of the next event is
\beqn
	f(t_2|t_1) = \lambda_\text{tot}\big(x(t_2)\big) 
	              \exp\!\Big[ -\int_{t_2}^{t_1} \lambda_\text{tot}\big(x(t)\big) \, \text{d}t \,\Big]. 
\eeqn
Hence, given that we have simulated the sweep from the end of the sweep to time $t_1$, the time $t_2$ of the next event is determined by solving the equation
\beqn
	\int_{t_2}^{t_1} \lambda_\text{tot}\big(x(t)\big) \, \text{d}t = \eta \label{eq:num_solve_rate}
\eeqn
numerically for $t_2$, where $\eta$ is an exponentially distributed variable with expected value unity. For some simple growth models it is possible to find explicit analytical expressions for $t_2$ as a function of $t_1$ and $\eta$; mostly, however, one must use numerical approximations of the integral. In this paper, we consider $x(t)$ in (\ref{eq:num_solve_rate}) to be a given, piecewise constant function. Also when we have explicit expressions for $x(t)$ it is convenient, and efficient, to take a number of samples at equally spaced points in time. We are then able to quickly find the interval containing the value of $t_2$ that solves (\ref{eq:num_solve_rate}) (if $x(t)$ is piecewise constant, the left-hand side of (\ref{eq:num_solve_rate}) is piecewise linear and continuous).

This concludes our review of the standard background coalescent.
There is only one problem with this picture: the rates $\lambda_\text{B}$ and $\lambda_\text{b}$ do not accurately describe the rate of birth-death events in the two sub-populations when we compare to simulations
using the 
Moran-model algorithm  described in section \ref{sec:moran}: 
we observe slight but statistically significant deviations for large 
values of $s$ (we find that the effect is negligible for $s < 0.03$, 
and is most significant  when both $s$ and $r$ are relatively large).

\begin{figure}
  \psfrag{xx}[t][]{$2Nx$}
  \psfrag{yy}[][t]{$\lambda_{\rm B}/2N-x$}
  \centerline{\includegraphics[width=8cm]{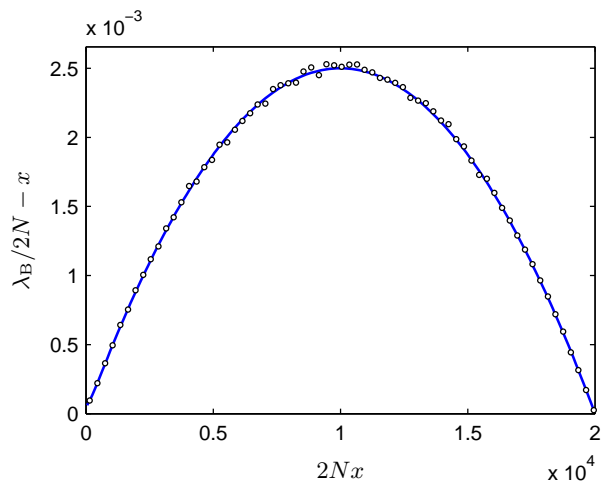}}
\caption{\label{fig:rate_events}
Shows the birth rate  of B alleles, $\lambda_{\rm B}$, as a function of $x$ for $N=10^4$, $s=0.01$, and $10^4$ Moran simulations 
(white circles). Also shown is the theory developed below (solid blue line). 
Note that the standard rates (\ref{eq:lambda_standard}) correspond to
$\lambda_{\rm B} = 2N x$.  }
\end{figure}

As is shown in Fig.~\ref{fig:rate_events}, the true birth 
rate of B alleles as a function of $x$ in the Moran model is given by the total rate of all events leading
to the birth of a B allele: combining eqs.~(\ref{eq:moran_rates}) and (\ref{eq:tilde_w}), we have  
\begin{align}\label{eq:lambda_B_true}
   \lambda_\text{B} 
    &= \widetilde{w}_{\text{B}\ra\text{B}} + \widetilde{w}_{\text{b}\ra\text{B}} \nonumber\\
    &= 2N \left[x + \frac{sx(1-x)}{1-\left(1-s\right)^{2Nx}} \right] .
\end{align}
Hence, the birth rate of B alleles is larger than expected from the standard model.
Since the total number of events is fixed at $(2N)^2$ per unit of time, the birth-rate of the b alleles is correspondingly smaller:
\begin{align}\label{eq:lambda_b_true}
     \lambda_\text{b} &= 2N -\lambda_\text{B}.
\end{align}
In general, we see that deviations from the standard rates are due to 
the difference in the birth rates of the two alleles. 
It is the selection process which causes extra births to happen in the B sub-population, and fewer births in the b sub-population.

\begin{figure}
\psfrag{x}[][]{$r$}
\psfrag{y}[][]{\emph{p2inb}}
	\includegraphics[width=8.5cm]{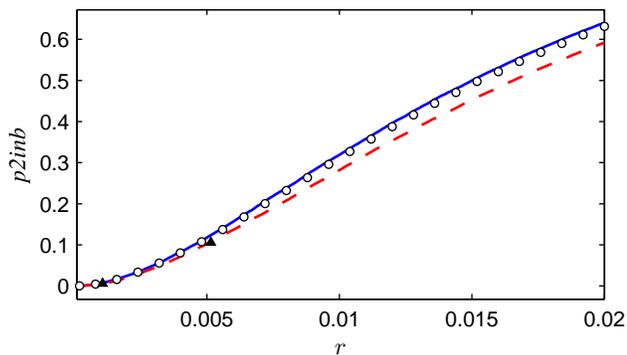}
\caption{\label{fig:moran_difference}
Probability \emph{p2inb} that two ancestral lines of a neutral locus
escape the sweep separately, as a function of the amount of recombination $r$ 
between the neutral and the selected locus. 
Shown are results of Moran-model simulations (circles), results of
the background coalescent with the growth $x(t)$ given by sampling the Moran process for the selected locus, using either the standard rates in the literature (\ref{eq:lambda_standard}), red dashed line, 
or the new rates (\ref{eq:lambda_B_true}) and (\ref{eq:lambda_b_true}), blue solid line. 
The coalescent simulations of \citet{durrett_schweinsberg04} (triangles) 
are consistent with the former, while our Moran model is much closer to the latter.
The parameters are: $N = 10^4$ and $s=0.1$. 
}
\end{figure}

In Fig.~\ref{fig:moran_difference} we illustrate the difference between choosing the birth-rates according to the standard method (\ref{eq:lambda_standard}), 
and according to (\ref{eq:lambda_B_true}), by measuring the probability \emph{p2inb} that two ancestral lines of a neutral locus escape the sweep separately. The parameters are 
$N=10^4$ and $s=0.01$, corresponding to moderately strong selection. The background coalescent using $\lambda_\text{B}$ from (\ref{eq:lambda_B_true}) is in good agreement with the Moran simulations, while the results using the rates (\ref{eq:lambda_standard}) exhibit a small but significant difference. 
Other quantities exhibit similar differences (not shown).

\section{Results and discussion}
\label{sec:results}

We have implemented the background coalescent for
a contiguous segment of neutral loci close to a selected
site (section \ref{sec:bg_coalescent}) using
the deterministic model $\bar{x}(t) = \bar{k}(t)/(2N)$
described in section \ref{sec:formulation_problem}:
$\bar{k}(t)$ is obtained by
solving $\langle \widetilde{t}_{{k}}\rangle = t$ for ${k}$,
as described in section \ref{sec:det_model}.

To establish the accuracy of  our algorithm, we compare its results to those of Moran-model simulations. In particular we 
compute the distribution over partitions at a neutral locus in the sample \citep[explained below 
in section B]{durrett_schweinsberg04}.

\subsection{Duration of the sweep}

\begin{figure}
  \psfrag{x}[t][]{$s$}
  \psfrag{y}[][t]{$\expt{\widetilde t_{2N}}/2N$}
	\includegraphics[width=8cm]{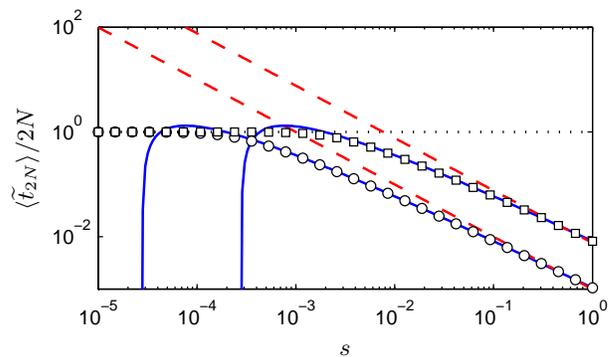}
\caption{\label{fig:sweep_time_approx}
Comparison of the exact expression (\ref{eq:expt_T_total}), symbols, for the expected duration of the sweep (in units of $2N$ generations) as a function of $s$, to the approximation (c.f. eq.~\ref{eq:expt_T_total_approx}, solid blue lines) and the logistic model (solid red lines), for $N = 10^3$ (squares) and $N=10^4$ (circles). As a reference, the result (36) is also shown (dotted line). }
\end{figure}

According to the results in section \ref{sec:expected_value_T_n}, 
we can use (\ref{eq:exp_T_n_sol}) to obtain a closed expression for $\expt{\widetilde{t}_{2N}}$, the expected duration of the sweep.
Because of symmetry, we can write $\expt{\widetilde t_{2N}}$ in the form
\beqn\label{eq:expt_T_total}
   \expt{\widetilde t_{2N}} = \sum_{k=1}^{2N-1}\frac{2\left(1-\omega^{2N-k}\right)\left(1-\omega^k\right)}{
   k \left(1-\omega\right)\left(1-\omega^{2N}\right)}.
\eeqn
In the limit $s \rightarrow 0$, we obtain the familiar result \citep[see, e.g.,][for a review]{ewens79}
\beqn
   \left.\expt{\widetilde t_{2N}}\right|_{s=0} = 2N - 1.
\eeqn
When $2Ns$ is large, we approximate $\omega^{2N} \approx 0$, and 
obtain to leading order
\beqn\label{eq:expt_T_total_approx}
    \expt{\widetilde t_{2N}} \approx 2\,\frac{\log(2Ns) + \gamma}{s}\,.
\eeqn 
Here $\gamma$ is Euler's constant, $\gamma \approx 0.577216$. This approximation is excellent: as is shown in Fig.~\ref{fig:sweep_time_approx}, the approximation breaks down only when $2Ns \lesssim 2$. Except for the $\gamma$-term, (\ref{eq:expt_T_total_approx}) is also the expected duration of the sweep one obtains in the 
diffusion approximation for the sweep conditioned on success \citep[Lemma 3.1]{etheridge_etal06}. 

This result should be contrasted with the deterministic logistic sweep, where the duration of the sweep is $2\log(2N-1)/s$. For large values of $s$, the duration is close to that of both the Moran model and to the approximation eq.~(\ref{eq:expt_T_total_approx}). Thus, quantities depending primarily on the duration of the sweep, such as the amount of recombination taking place during the sweep, will be accurately described in the logistic model when the selection is strong. From (\ref{eq:expt_T_total_approx}), and in Fig.~\ref{fig:sweep_time_approx}, we see that this happens when $|\log(s)|$ is small compared to $\log(2N)$. 
When $s$ is small, however, the duration of the sweep in the logistic model is very different from that of the Moran model, and consequently we expect a clear difference in the effect of the sweep on the neutral loci nearby.

\subsection{Partitions}

\begin{figure}
\psfrag{x}[][]{$r$}
\psfrag{p2cinB}[][]{ \emph{p2ciB}}
\psfrag{p2inb}[][]{ \emph{p2inb}}
\psfrag{p2cinb}[][]{ \emph{p2cinb}}
\psfrag{p1B1b}[][]{ \emph{p1B1b}}
	\includegraphics[width=8.5cm]{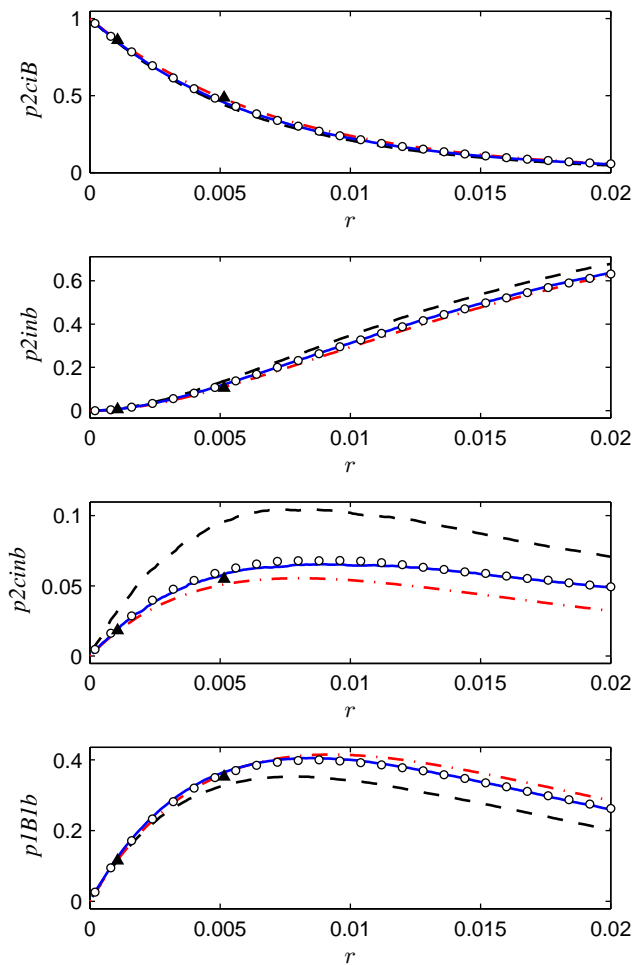}
\caption{\label{fig:part_distr_a}
The distribution over the partitions as a function of the genetic distance $r$ from the selected locus. We show one panel for each of the four partitions.
The population size is $N = 10^4$, and the selection parameter $s = 0.1$.  
The data shown are: Moran simulations (circles), logistic model (dashed black
line), our own  model (solid blue line), the DS-algorithm (dash-dotted red line), and
coalescent simulations of \citet{durrett_schweinsberg04} (triangles). 
}
\end{figure}

\begin{figure}
\psfrag{x}[][]{$r$}
\psfrag{p2cinB}[][]{ \emph{p2ciB}}
\psfrag{p2inb}[][]{ \emph{p2inb}}
\psfrag{p2cinb}[][]{ \emph{p2cinb}}
\psfrag{p1B1b}[][]{ \emph{p1B1b}}
	\includegraphics[width=8.5cm]{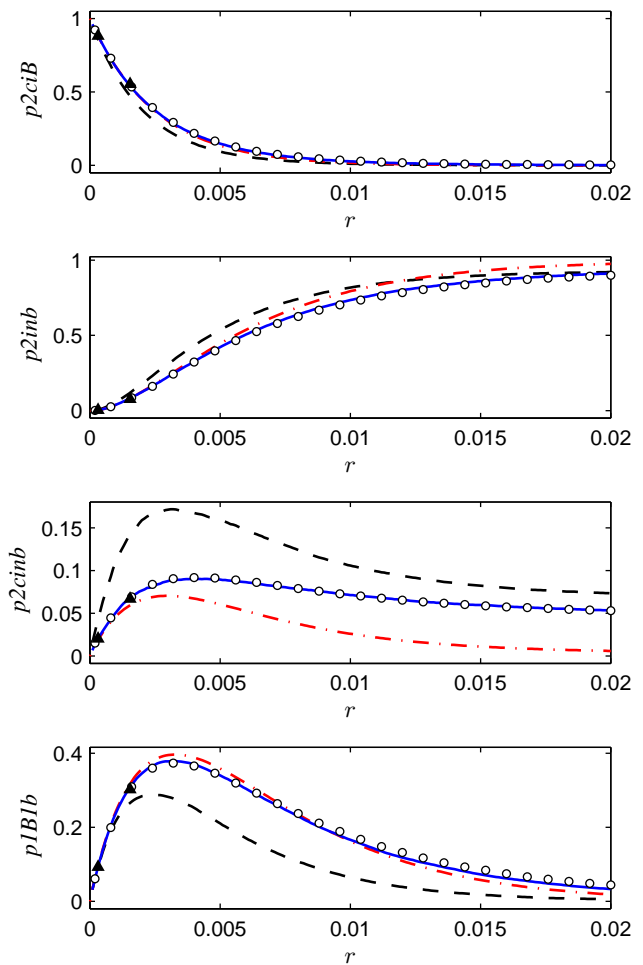}
\caption{\label{fig:part_distr_b}
As Fig. \ref{fig:part_distr_a}, but for $s = 0.03$.  
}
\end{figure}

\begin{figure}
\psfrag{x}[][]{$r$}
\psfrag{p2cinB}[][]{ \emph{p2ciB}}
\psfrag{p2inb}[][]{ \emph{p2inb}}
\psfrag{p2cinb}[][]{ \emph{p2cinb}}
\psfrag{p1B1b}[][]{ \emph{p1B1b}}
	\includegraphics[width=8.5cm]{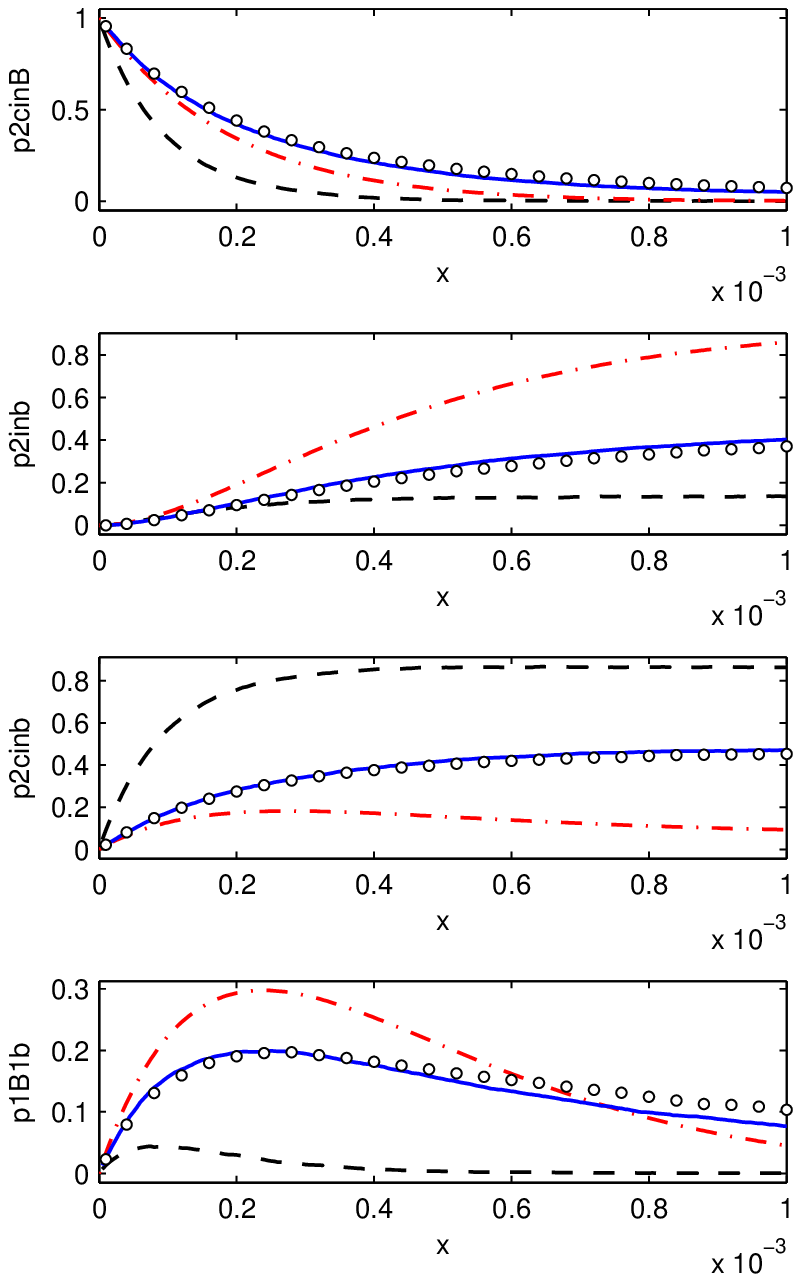}
\caption{\label{fig:part_distr_c}
As Fig. \ref{fig:part_distr_a}, but for $s = 0.001$. 
}
\end{figure}

In this subsection,  we consider the distribution of partitions at a neutral locus at distance $r$ from the selected locus in a sample of two individuals in the population. The partitions are defined as follows \citep{donelly86, durrett_schweinsberg04}. Suppose we follow the ancestral lines of the neutral locus in the two individuals back in time through the sweep. Because of recombination, the lines may move from the B population to the b population, and (with a rather small probability) back again. They may coalesce in one of the populations, or stay separate during the whole sweep. For two lines, we have four distinct cases: both lines coalesce during the sweep and the resulting line is trapped by the sweep (the probability
for this to happen is denoted by \emph{p2cinB}); one line escapes the sweep and the other is trapped (\emph{p1B1b}); both lines escape the sweep but do not coalesce (\emph{p2inb}); the lines coalesce and then escape, or escape separately and then coalesce (much less likely), denoted by \emph{p2cinb}.

Far away from the sweep, one expects all lines to escape independently. For large population sizes it is unlikely that 
lines coalesce during the sweep, but it becomes more common when the population size is relatively low (e.g., for $N \sim 10^3$). Close to the selected locus, nearly all lines are trapped in the sweep. The frequency of the case where one line is trapped and the other line escapes has a maximum for intermediate 
genetic distances $r$. 

In Fig.~\ref{fig:part_distr_a} we compare the four models: 
the Moran model, the logistic-sweep model, the DS-algorithm, and our own algorithm, when $N=10^4$ and $s=0.1$, corresponding to strong selection. Also shown are the coalescent simulations
of \citet{durrett_schweinsberg04}. 
The plot covers the approximate range of validity quoted
by \citet{durrett_schweinsberg04} for their algorithm: $r \lesssim s/\ln 2N$ which evaluates
to $\approx 0.01$. Over this range, all curves except the logistic model agree.
In particular, the logistic model gives a higher value for \emph{p2cinb} than expected; the most likely reason for this deviation is that the duration of the sweep is slightly too long in the logistic model (c.f. Fig.~\ref{fig:sweep_time_approx}). 

Figs.~\ref{fig:part_distr_b} and \ref{fig:part_distr_c} show the same quantities as Fig. \ref{fig:part_distr_a}
but for $s=0.03$ and $s=0.001$, respectively. 
The range of validity of the DS-algorithm is 
$r < s/\ln 2N$ which is $0.003$ in Fig.~\ref{fig:part_distr_b}, 
and $10^{-4}$ in Fig.~\ref{fig:part_distr_c}. Within
this range, all curves except the logistic model agree approximately.

For larger values of $r$, 
the most important contribution to the difference between the Moran model and the DS-algorithm is that the latter ignores recombination events and coalescent events during the middle and late stages of the sweep. As can be seen in the figures, this is a very good approximation provided $r$ is sufficiently small, or provided
the sweep is sufficiently short. The accuracy of the logistic model quickly deterioates
as $s$ decreases. Again, the most important reason is that the sweep is too long compared to the Moran model. 

Our algorithm, by contrast works well also for large values of $r$ and small values of $s$, although it is clear that the deviations from the Moran model become larger for  smaller values of $s$. 
This  is to be expected since the fluctuations of the sweep increase with decreasing $s$.

Last but not least we emphasize that the curves in Figs.~\ref{fig:part_distr_a}--\ref{fig:part_distr_c}
are obtained by a single run of our program for a contiguous stretch of DNA adjacent to the selected site.
The DS-algorithm requires a separate simulation for each value of $r$.

\section{Conclusions}
\label{sec:conclusions}

We have implemented a new model for
genetic hitch-hiking based on a deterministic
approximation for the growth of the favoured-allele
frequency during the selective sweep, 
in combination with a coalescent process for  a locus (or set of loci)
close to the selected locus.  By comparison with direct Moran-model simulations
we could show that our new model is very accurate. Two reasons for this
success are: our model faithfully approximates the expected duration
of the selective sweep, and it is conditioned on the success of the
sweep. 

Our algorithm is as easily implemented as the standard logistic model,
but is far more accurate, even applicable beyond the range 
of parameters given by \citet{durrett_schweinsberg04} for their
algorithm. For practical purposes it is important that 
the sweep is not assumed to happen instantaneously, so mutations 
occuring during the sweep are not neglected. Furthermore,
the algorithm determines the fate of a contiguous segment
of neutral loci in the vicinity of the selected locus.
Figs. 8-10, for example, were obtained by one single run
of our algorithm.

Our results have implications beyond the immediate context of this
article. First, we introduced a new approximate representation
of selective sweeps (the \lq sorted' sweep) which locally averages
over fluctuations in the favoured-allele frequency.  We suspect
that this approximation retains the fluctuations relevant
for an accurate description of the genealogies of neutral loci close to
the selected site. In which range of parameters this is true
will be the subject of a subsequent study.
Second, in the coalescent for the neutral loci, we have shown
that the standard expression for the rates (\ref{eq:lambda_standard})
must be modified.
We expect that similar modifications are necessary in other cases,
e.g. Moran models with changing population-sizes, as for instance
in population expansions and bottlenecks.

We conclude with describing a possible application for our model.
It will be of use in efficiently and accurately determining
log-likelihood surfaces for the parameters $s$ and $N$ in the Moran model of directional selection \citep[see, e.g.,][]{coop_griffiths04} where an accurate and computationally
efficient model is required.
We believe that our deterministic approximation will be of use in this context.

\subsection*{Acknowledgments}
BM acknowledges support from Vetenskapsr\aa{}det.

\bibliography{genetics}

\begin{thebibliography}{25}
\expandafter\ifx\csname natexlab\endcsname\relax\def\natexlab#1{#1}\fi

\bibitem[{{\sc Barton}(1998)}]{barton98}
{\sc Barton, N.~H.}, 1998 The effect of hitch-hiking on neutral genealogies.
\newblock Genet. Res. Camb. {\bf 72}: 123--133.

\bibitem[{{\sc Braverman} {\em et~al.\/}(1995){\sc Braverman}, {\sc Hudson},
  {\sc Kaplan}, {\sc Langley} and {\sc Stephan}}]{bra95:hit}
{\sc Braverman, J.~M.}, {\sc R.~R. Hudson}, {\sc N.~L. Kaplan}, {\sc C.~H.
  Langley}, and {\sc W.~Stephan}, 1995 The hitchhiking effect on the site
  frequency spectrum of {DNA} polymorphism.
\newblock Genetics {\bf 140}: 783--796.

\bibitem[{{\sc Coop} and {\sc Griffiths}(2004)}]{coop_griffiths04}
{\sc Coop, G.}, and {\sc R.~C. Griffiths}, 2004 Ancestral inference on gene
  trees under selection.
\newblock Theor. Popul. Biol. {\bf 66}: 219--232.

\bibitem[{{\sc Donnelly}(1986)}]{donelly86}
{\sc Donnelly, P.}, 1986 Partition structures, {P}olya urns, the {E}wens
  sampling formula and the ages of alleles.
\newblock Theor. Pop. Biol. {\bf 30}: 271--288.

\bibitem[{{\sc Durrett}(2002)}]{durrett02}
{\sc Durrett, R.}, 2002 {\em Probability Models for DNA Sequence Evolution\/}.
\newblock Springer, New York.

\bibitem[{{\sc Durrett} and {\sc Schweinsberg}(2004)}]{durrett_schweinsberg04}
{\sc Durrett, R.}, and {\sc J.~Schweinsberg}, 2004 Approximating selective
  sweeps.
\newblock Theor. Popul. Biol. {\bf 66}: 129 -- 138.

\bibitem[{{\sc Etheridge} {\em et~al.\/}(2006){\sc Etheridge}, {\sc
  Pfaffelhuber} and {\sc Wakolbinger}}]{etheridge_etal06}
{\sc Etheridge, A.}, {\sc P.~Pfaffelhuber}, and {\sc A.~Wakolbinger}, 2006 An
  approximate sampling formula under genetic hitchhiking.
\newblock Annals of Applied Probability {\bf 16}: 685--729.

\bibitem[{{\sc Ewens}(1979)}]{ewens79}
{\sc Ewens, W.~J.}, 1979 {\em Mathematical population genetics\/}.
\newblock Springer, Berlin.

\bibitem[{{\sc Fisher}(1930/1999)}]{fis30:gen}
{\sc Fisher, R.~A.}, 1930/1999 {\em The Genetical Theory of Natural
  Selection\/}.
\newblock Oxford University Press, variorum edition.

\bibitem[{{\sc Hudson}(1983)}]{hudson83}
{\sc Hudson, R.~R.}, 1983 Properties of a neutral allele model with
  intragenetic recombination.
\newblock Theor. Pop. Biol. {\bf 23}: 183--201.

\bibitem[{{\sc Hudson}(1990)}]{hudson90}
{\sc Hudson, R.~R.}, 1990 Gene genealogies and the coalescent process.
\newblock In D.~Futuyma and J.~Antonovics, editors, {\em Oxford Surveys in
  Evolutionary Biology\/}. Oxford University Press, Oxford, 1--43.

\bibitem[{{\sc Hudson}(2002)}]{hudson02}
{\sc Hudson, R.~R.}, 2002 Generating samples under a {W}right-{F}isher neutral
  model of genetic variation.
\newblock Bioinformatics {\bf 18}: 227--338.

\bibitem[{{\sc Kaplan} {\em et~al.\/}(1988){\sc Kaplan}, {\sc Darden} and {\sc
  Hudson}}]{kap88:coa}
{\sc Kaplan, N.~L.}, {\sc T.~Darden}, and {\sc R.~R. Hudson}, 1988 The
  coalescent process in models with selection.
\newblock Genetics {\bf 120}: 819--829.

\bibitem[{{\sc Kaplan} {\em et~al.\/}(1989){\sc Kaplan}, {\sc Hudson} and {\sc
  Langley}}]{kap89:hit}
{\sc Kaplan, N.~L.}, {\sc R.~R. Hudson}, and {\sc C.~H. Langley}, 1989 The
  ``hitchhiking effect" revisited.
\newblock Genetics {\bf 123}: 887--899.

\bibitem[{{\sc Kim} and {\sc Stephan}(2002)}]{kim_stephan02}
{\sc Kim, Y.}, and {\sc W.~Stephan}, 2002 Detecting a local signature of
  genetic hitchhiking along a recombining chromosome.
\newblock Genetics {\bf 160}: 765--777.

\bibitem[{{\sc Kingman}(1982)}]{kin82:coa}
{\sc Kingman, J. F.~C.}, 1982 The coalescent.
\newblock Stochastic Processes and their Applications {\bf 13}: 235--248.

\bibitem[{{\sc Maynard~Smith} and {\sc Haigh}(1974)}]{may74:hit}
{\sc Maynard~Smith, J.}, and {\sc J.~Haigh}, 1974 The hitch-hiking effect of a
  favourable gene.
\newblock Genetical Research, Cambridge {\bf 23}: 23--35.

\bibitem[{{\sc McPeek} and {\sc Speed}(1995)}]{mcpeek_speed95}
{\sc McPeek, M.~S.}, and {\sc T.~P. Speed}, 1995 Modelling interference in
  genetic recombination.
\newblock Genetics {\bf 139}: 1031--1044.

\bibitem[{{\sc Moran}(1958)}]{moran58}
{\sc Moran, P. A.~P.}, 1958 Random processes in genetics.
\newblock Proc. Cambridge Philos. Soc. {\bf 54}: 60--71.

\bibitem[{{\sc Nordborg}(2001)}]{nor01:coa}
{\sc Nordborg, M.}, 2001 Coalescent theory.
\newblock In D.~J. Balding, M.~Bishop and C.~Cannings, editors, {\em Handbook
  of Statistical Genetics\/}, chapter~7. John Wiley \& Sons, 179--212.

\bibitem[{{\sc Otto} and {\sc Barton}(1997)}]{per97:evo}
{\sc Otto, P.~S.}, and {\sc N.~H. Barton}, 1997 The evolution of recombination:
  removing the limits to natural selection.
\newblock Genetics {\bf 147}: 879--906.

\bibitem[{{\sc Przeworski}(2002)}]{przeworski02}
{\sc Przeworski, M.}, 2002 The signature of positive selection at randomly
  chosen loci.
\newblock Genetics {\bf 160}: 1179 -- 1189.

\bibitem[{{\sc Schweinsberg} and {\sc Durrett}(2005)}]{schweinsberg_durrett05}
{\sc Schweinsberg, J.}, and {\sc R.~Durrett}, 2005 Random partitions
  approximating the coalescence of lineages during a selective sweep.
\newblock Preprint (May 13) .

\bibitem[{{\sc Stephan} {\em et~al.\/}(1992){\sc Stephan}, {\sc Wiehe} and {\sc
  Lenz}}]{stephan_etal92}
{\sc Stephan, W.}, {\sc T.~Wiehe}, and {\sc M.~W. Lenz}, 1992 The effect of
  strongly selected substitutions of neural polymorphisms: Analytical results
  based on diffusion theory.
\newblock Theor. Pop. Biol. {\bf 41}: 237--254.

\bibitem[{{\sc Wright}(1931)}]{wri31:evo}
{\sc Wright, S.}, 1931 Evolution in {M}endelian populations.
\newblock Genetics {\bf 16}: 97--159.

\end{thebibliography}

\end{document}